\documentclass[useamsfonts,dvipdfmx]{pasj00}
\newcommand{\ion}[2]{#1$\,${\sc #2}}
\newcommand{\HI}{\ensuremath{\mbox{\ion{H}{i}}}} 
\newcommand{\HII}{\mbox{\ion{H}{ii}}} 

\begin{document}
\SetRunningHead{T. Okamoto et al.}{Evolution of galaxy population}

\title{Reproducing cosmic evolution of galaxy population from $z = 4$ to $0$} 
\author{Takashi \textsc{Okamoto}$^1$ %
}
\affil{$^1$Department of Cosmosciences, Graduates School of Science, Hokkaido University, N10 W8, Kitaku, Sapporo, 060-0810, Japan}
\email{okamoto@astro1.sci.hokudai.ac.jp}

\author{Ikkoh \textsc{Shimizu}$^2$}
\affil{$^2$Department of Astronomy, School of Science, The University of Tokyo, 7-3-1 Hongo, Bunkyo-ku, Tokyo 113-0033}
\email{shimizu@astron.s.u-tokyo.ac.jp}
\and
\author{Naoki {\sc Yoshida}$^{3, 4}$}
\affil{$^3$Department of Physics, The University of Tokyo, 7-3-1 Hongo, Bunkyo-ku, Tokyo 113-0033, Japan\\%
$^4$Kavli Institute for the Physics and Mathematics of the Universe, TODIAS, %
The University of Tokyo, 5-1-5 Kashiwanoha, Kashiwa, Chiba 277-8583, Japan}\email{naoki.yoshida@phys.s.u-tokyo.ac.jp}

%


\maketitle

\begin{abstract}
  We present cosmological hydrodynamic simulations performed 
  to study evolution of galaxy population. 
  The simulations follow timed release of mass, energy, and 
  metals by stellar evolution and employ phenomenological 
  treatments of supernova feedback, pre-supernova feedback 
  modeled as feedback by radiation pressure from massive stars, 
  and quenching of gas cooling in large halos. 
  We construct the fiducial model so that it reproduces the 
  observationally estimated galaxy stellar mass functions and the 
  relationships between the galaxy stellar mass and the host halo mass 
  from $z = 4$ to 0. 
  We find that the fiducial model constructed this way naturally 
  explains the cosmic star formation history, the galaxy downsizing, 
  and the star formation rate and metallicity of the 
  star-forming galaxies.  
  The simulations without the quenching of the gas cooling in 
  large halos overproduce massive galaxies at $z < 2$ and fail to 
  reproduce galaxy downsizing. 
  The simulations that do not employ the radiation pressure feedback 
  from young stars predict too strong redshift evolution of the 
  mass-metallicity relation. Furthermore, the slope of the relation 
  becomes too steep at low redshift without the radiation pressure 
  feedback. 
  The metallicity dependence in the radiation 
  pressure feedback is a key to explain the observed mass-metallicity 
  relation. 
  These facts indicate that these two processes in addition to 
  supernova feedback are essential for galaxy evolution.  
  Our simple phenomenological model is suitable to  
  construct a mock galaxy sample to study physical properties 
  of observed galaxy populations. 
\end{abstract}

\section{Introduction}

Understanding galaxy formation is a challenging problem whose solution will 
require a concerted approach combining observational and theoretical work. 
There have been substantial advances on both fronts in the past decades. 
Numerical simulations are a powerful theoretical tool to study cosmic 
structure formation. 
$N$-body simulations are now able to predict non-linear growth of the dark 
matter-dominated density perturbations in great detail 
\citep{millennium, aquarius, ViaLacteaII, boylan-kolchin09}. 
Consequently, gravitational assembly of structure in a $\Lambda$-dominated 
Cold Dark Matter ($\Lambda$CDM) Universe is well understood and mostly 
consistent with observations. 
In order to make a direct comparison with observations,   
simulations must involve luminous matters (baryons) besides dark matter 
and dark energy. 
While semi-analytic models (e.g. \cite{WF91}; \cite{kwg93};  
\cite{sp99}; \cite{clbf00}; \cite{on04}; \cite{nuGC})  
can paint galaxies onto dark matter distribution, 
hydrodynamic simulations can directly explore evolution of the galaxy 
population and the intergalactic medium (IGM) simultaneously and 
self-consistently (e.g. \cite{khw92}; \cite{whk97}; \cite{keres05}; 
\cite{od06}; \cite{ocvirk08}; \cite{gimic}; \cite{owls}; 
\cite{vogelsberger13}). 
The baryonic processes that are essential for galaxy formation,
such as gas cooling, star formation, and stellar and active galactic 
nuclei (AGN) feedback, constitute a complicated and highly non-linear 
network. 
Modeling them appropriately in hydrodynamic simulations is hence 
the major challenge for the theoretical studies of galaxy formation 
(see \cite{oka05}; \cite{Aquila}). 

Some recent simulations successfully produce realistic galaxies 
(\cite{ofjt10, gov10, eris, okamoto13, magicc, marinacci13}). 
The key ingredient is undoubtedly stellar feedback such as 
supernova (SN) feedback that ejects 
gas from galaxies to prevent too efficient star formation. 
While it is possible to drive winds by resolving the detailed 
structure of the interstellar medium (ISM) in very high resolution 
simulations \citep{hopkins12a}, most of cosmological 
simulations invoke phenomenological treatments of feedback 
because of limited numerical resolution. 
Some studies employ explicit winds, either hydrodynamically decoupled 
\citep{sh03} or coupled \citep{ds08}. The wind properties may depend 
on the galaxy properties \citep{od06, ofjt10}. 
Other popular way of implementing effective feedback is injecting 
thermal energy into the ISM and then shutting off cooling of heated gas
for a while \citep{tc01, sti06, magicc}. Adding the feedback energy as 
non-thermal energy (e.g. turbulence) which decays in time-scale 
much longer than the cooling time has a similar effect 
\citep{teyssier13}. 

We have applied the feedback model developed by \citet{onb08} 
to large scale simulations to study high redshift galaxy 
populations, such as Lyman-$\alpha$ emitters \citep{shimizu11}, 
sub-mm galaxies \citep{shimizu12}, and Lyman break galaxies at 
$z > 7$ \citep{shimizu14, inoue14}. 
While these simulations reproduce many observed properties of 
high redshift galaxies, our studies utilizing the large scale 
simulations have been limited to high redshift because too 
massive galaxies form at low redshift (see \cite{shimizu12}).  
We also note that too many stars form in low mass halos at 
high redshift if we normalize the model to reproduce 
the luminosity (or stellar mass) function of the local faint 
galaxies \citep{moster13, okamoto13}. 

The first problem is well-known; stellar feedback alone cannot 
prevent monster galaxies from forming and hence we need a physical 
process that operates preferentially in large halos to quench  
gas cooling there (e.g. \cite{ben03}). 
The top candidate of such a process is so-called AGN {\it radio mode} 
feedback \citep{croton06, bower06}. 
\citet{sijacki07} and \citet{onb08} suggest that this radio mode feedback 
is naturally realized by considering the change of the accretion modes 
onto a supermassive blackhole. In fact, simulations including this 
feedback roughly reproduce galaxy stellar mass functions and stellar mass 
to halo mass relations for massive galaxies 
\citep{vogelsberger13, torrey14}. 

A remedy for the second problem has been recently identified by \citet{magicc}; 
feedback prior to an SN, such as stellar winds and radiation from massive 
stars, is needed to match stellar mass-halo mass relations over a wide 
redshift range \citep{magicc, kannan13, aumer13}.

The aim of this paper is to update our galaxy formation model originally 
developed by \citet{onb08} and  \citet{ofjt10} 
by adding several new feedback processes so that we can apply it for 
studies of galaxy population over wider mass and redshift ranges. 
We test our models against various observations and reveal roles of 
each feedback process to present a fiducial model.  

The paper is organized as follows. In section 2, we describe our 
simulations and provide descriptions of our modeling of baryonic 
processes. We present our results at $0 \le z \le 4$ and compare 
them with the available observational estimates in section 3.  
We investigate resolution effects in section 4. 
Finally, we summarize our results and discuss future 
applications of the new model in section 5.  


\section{Simulations}

We first describe our cosmological hydrodynamic simulations. 
The simulation code is based on an early version of Tree-PM 
smoothed particle hydrodynamics (SPH) code {\scriptsize GADGET-3} 
which is a successor of Tree-PM SPH code {\scriptsize GADGET-2} 
\citep{spr05}. We have implemented a time-step limiter \citep{sm09} 
that reduces the time-step of a gas particle if it is too long 
compared to the neighboring gas particles. 
We have also added an artificial conductivity \citep{rosswog07, price08}, 
in order to capture the instabilities at contact surfaces 
(see also \cite{kawata13}), 
which cannot be captured by the standard SPH \citep{oka03, agertz07}, 
together with the time dependent artificial viscosity \citep{mm97}.

We assume $\Lambda$CDM cosmology with the following parameters: 
$\Omega_{\rm 0} = 0.318$, $\Omega_\Lambda = 0.682$, 
$\Omega_{\rm b} = 0.049$, $\sigma_8 = 0.835$, $n_{\rm s} = 0.962$, 
and a Hubble constant of $H_0 = 100~h~{\rm km}~{\rm s^{-1}}$~Mpc$^{-1}$, 
where $h = 0.67$. These parameters are consistent with Planck 2013 
results \citep{planck}.  
Throughout this paper, we use a cosmological periodic box of the side 
length of 40~$h^{-1}$~Mpc and start simulations from $z = 127$ 
unless otherwise stated. 
We employ $256^3$ dark matter particles and the same number of 
SPH particles. The total number the particles can change owing to 
the star formation. 
The mass of a dark matter particle is $2.9 \times 10^8 h^{-1} M_\odot$ 
and that of an SPH particle is $5.2 \times 10^7 \ h^{-1} M_\odot$.  
The gravitational softening length is set to $8.6 \ h^{-1}$~kpc in comoving 
units both for the dark matter and SPH particles (and stars) 
until $z = 3$; thereafter, it is frozen in physical units at 
the value, $2.2 h^{-1}$~kpc.

\subsection{Baryonic processes}

The simulations include many physical processes that are relevant 
to galaxy formation.  
Both photo-heating by a spatially uniform, time-evolving 
ultra-violet background and radiative cooling depend on 
gas metallicity as described in \citet{wss09}. 
The cooling and heating rates are computed individually for 
eleven elements (H, He, C, N, O, Ne, Mg, Si, S, Ca, and Fe). 
We track only nine elements and we take S and Ca to be 
proportional to Si \citep{wie09b}. 
As in \citet{oka05}, we use the smoothed metallicity instead 
of the particle metallicity to compute the photo-heating and 
radiative cooling rates and to give the initial amount of 
each individual chemical element to newly born stars. 
Note however that we do not include explicit metal diffusion. 

We assume star formation takes place when gas density exceeds 
a threshold density for star formation 
($n_{\rm H} > n_{\rm H, th} = 0.1~{\rm cc}^{-1}$), 
gas temperature is sufficiently low 
($T < T_{\rm th} = 15000~{\rm K}$), and flows are 
converging ($\nabla \cdot \boldsymbol{v} < 0$). 

While, in our previous studies, we have treated a star forming 
gas particle as a hybrid particle that contains hot and cold phases 
\citep{onb08, ofjt10, shimizu11, shimizu12, shimizu14}, 
we abandon doing this in this paper since global star 
formation properties of simulated galaxies are highly 
self-regulated by strong feedback and are almost 
independent of the treatment of the equation 
of states of the star-forming gas \citep{ofjt10, owls}. 

The star formation rate (SFR) density, $\dot{\rho}_*$, 
for a gas particle with the density, $\rho$, is then simply given by 
\begin{equation}
  \dot{\rho}_* = c_* \frac{\rho}{t_\mathrm{dyn}}, 
\end{equation}
where $c_*$ and $t_\mathrm{dyn}$ are respectively 
the dimensionless star formation 
efficiency parameter and the local dynamical time. 
This formula corresponds to the Schmidt law that implies 
an SFR density proportional to $\rho^{1.5}$.
We set $c_* = 0.01$ in order to reproduce the observed 
relation between the surface gas density and the surface star 
formation rate density \citep{ken98}. 
Technically, we compute the probability, ${\cal P}_*$, for an SPH 
particle with which it spawns a new star particle of mass, $m_*$, during 
a time-step, $\Delta t$, as 
\begin{equation}
  {\cal P}_* = \frac{m_{\rm SPH}}{m_*} 
    \left[1 - \exp\left(-\frac{\Delta t}{t_*}\right)\right], 
\end{equation}
where $m_{\rm SPH}$ is the mass of the gas particle and the 
star formation time-scale $t_*$ is defined as 
$t_* = t_{\rm dyn}/c_*$. 
We use $m_* = m_{\rm SPH}^{\rm orig}/2$ throughout this paper, 
where $m_{\rm SPH}^{\rm orig}$ denotes the original SPH 
particle mass. 

Stellar evolution is modeled as in \citet{ofjt10}. 
We employ the Chabrier initial mass function 
(IMF: \cite{chabrier03}) and we use metallicity dependent 
stellar lifetimes and chemical yields \citep{pcb98, marigo01}. 
The production of metals by SNe and AGB stars, stellar mass 
loss, and stellar feedback all take place on the time-scale 
dictated by stellar evolution considerations. 
When we compare our results with observational estimates that
assume different IMFs, we convert them to the Chabrier 
IMF\footnote{We convert stellar mass and SFR obtained by assuming the 
Salpeter IMF \citep{salpeter} and the Kroupa IMF \citep{kroupa01} into 
those with the Chabrier IMF by dividing them by 1.8 and 1.1, 
respectively.}.  

We now discuss the numerical implementation of important 
{\it subgrid} processes included in our simulations. 

\subsection{Stellar feedback}

In our subgrid model, stellar feedback gives rise to 
a wind by imparting kinetic energy and momentum to 
nearby gas particles. The wind is characterized by 
its initial speed, $v_{\rm w}$. 

The direction of the wind is chosen at random to be either parallel 
or antiparallel to the vector 
$(\boldsymbol{v} - \bar{\boldsymbol{v}}) \times \boldsymbol{a}_{\rm grav}$, 
where $\bar{\boldsymbol{v}}$ is the average velocity of 
the neighboring 64 dark matter particles and $\boldsymbol{a}_{\rm grav}$ 
is the gravitational acceleration. \citet{ofjt10} found that 
the direction of the wind particle velocity defined in this way leads to 
wind particles being ejected preferentially along the rotation 
axis of a spinning object, and thus generating an `axial' wind 
\citep{sh03}. 
We assume that the newly launched wind particles are decoupled from 
hydrodynamic interactions for a brief period of time \citep{sh03}. 
Full hydrodynamic interactions are enabled once a wind particle 
leaves the star-forming region ($n_{\rm H} < 0.01~{\rm cc}^{-1}$) 
or after the time, ($10~{\rm kpc})/v_{\rm w}$, has elapsed, 
whichever occurs earlier. 

\vspace{0.1cm}

\subsubsection{Supernova-driven winds}

\vspace{0.1cm}

We assume that some fraction of the energy released from 
SNe is potentially available to power the kinetic energy 
of the wind. 
Each SN releases energy of $10^{51}$~erg and the fraction of 
energy used to drive wind is controlled by an efficiency 
parameter, $\eta_{\rm SN}$. 
As in \citet{parry12} and \citet{okamoto13}, 
we suppose that only Type II SNe are responsible for 
driving winds and the energy released from Type Ia SNe is 
distributed as thermal energy to surrounding gas particles. 

During any given time-step, a gas particle may receive 
supernova energy, $\Delta Q$, from one or more neighboring 
star particles. 
If this happens, the particle is given a probability, 
${\cal P}_{\rm w}^{\rm SN}$, with which it becomes a wind particle during 
that time-step: 
\begin{equation}
  {\cal P}_{\rm w}^{\rm SN} 
  = \frac{\Delta Q}{\onehalf m_{\rm SPH} v_{\rm w, SN}^2},  
  \label{eq:energy-driven}
\end{equation}
where $v_{\rm w, SN}$ is the initial wind speed. 
The value of $v_{\rm w, SN}$ is defined to be proportional to the 
one-dimensional velocity dispersion of the neighboring 
dark matter particles, $\sigma$, namely, 
\begin{equation}
  v_{\rm w, SN} = \kappa_{\rm w}^{\rm SN} \sigma, 
\end{equation}
where $\kappa_{\rm w}^{\rm SN}$ is a proportional constant that defines 
the wind speed with respect to the local velocity dispersion. 
This scaling is motivated by the data on galactic outflows 
\citep{mar05}. 
\citet{ofjt10} have shown that doing this reproduces the observed luminosity 
function and the luminosity-metallicity relation of the Local Group 
satellite galaxies. \citet{puchwein13} have reported a good match to the 
faint end slope of the observed galaxy stellar mass function 
with this scaling. 

\vspace{0.1cm}

\subsubsection{Pre-supernova feedback}

\vspace{0.1cm}

\citet{magicc} identified that the lag between a star formation 
event and an SN explosion ($\sim 10~{\rm Myr}$) is likely to 
be a source of the difficulty for simulations in reproducing 
the observationally suggested relation between stellar mass 
and halo mass. They reproduced the relation by distributing 
10~\% of the bolometric luminosity emitted by young stars to 
the surrounding star-forming gas as thermal energy over $0.8$~Myr 
time period before an SN goes off. 

Recently some numerical studies have indicated the importance 
of radiation pressure feedback from young stars on galaxy 
evolution \citep{hopkins11, hopkins12b, chattopadhyay12, wise12}. 
Momentum injection by stellar winds also cannot be ignored 
\citep{hopkins12b}. Massive stars can radiatively drive stellar 
winds from their envelopes during the first stage of 
evolution, reaching terminal velocities of 1000--3000~km~s$^{-1}$.
The momentum injection by the stellar winds is roughly the same order as 
that by the radiation pressure \citep{agertz13}. 
We thus consider the momentum driven winds 
\citep{od06, od08} as the pre-SN feedback in this paper.  

The momentum injection rate from radiation can be written as 
\begin{equation}
  \dot{p}_{\rm rad} = (\eta_1 + \eta_2 \tau_{\rm IR}) 
  \frac{L(t)}{c}, 
  \label{eq:radiation}
\end{equation}
where $\eta_1$ and $\eta_2$ respectively define the momentum 
transfer efficiencies for the direct absorption/scattering and 
for the multi-scattering by infrared photons re-radiated by dust 
particles; $\tau_{\rm IR}$ is the infrared optical depth, $L(t)$ 
is the luminosity of the stellar population, which we calculate 
using a population synthesis code {\scriptsize P\'EGASE} 
\citep{pegase} as a function of the age and the metallicity 
of the star particle, and $c$ is the speed of light. 

The first term of equation~(\ref{eq:radiation}) should be 
proportional to $1 - \exp(-\tau_{\rm UV})$, where $\tau_{\rm UV}$ 
is the ultraviolet (UV) optical depth.   
Since the dust and \HI \ opacities in the UV present in 
star-forming regions are very large, it is reasonable to assume 
that the value of $\eta_1$ is around unity. 
The estimation of $\tau_{\rm IR}$ is problematic because we cannot 
resolve star-forming clouds in large volume simulations. 
We here simply assume that $\tau_{\rm IR}$ scales linearly with 
the metallicity of a star particle\footnote{We must use the IR optical 
depth of a cloud in which the stellar population (i.e. star particle) 
is embedded. The metallicity of the cloud should be comparable to that of 
the stellar population because the stellar population inherits the metallicity 
of the parent cloud.} (see \cite{aumer13} and \cite{agertz13} 
for more elaborate estimations of $\tau_{\rm IR}$). 
Under this assumption, equation~(\ref{eq:radiation}) reduces to 
\begin{equation}
  \dot{p}_{\rm rad}(Z, t) = \left[\eta_1 + \tau_0 \left(\frac{Z}{Z_\odot}
  \right)\right] \frac{L(t)}{c}, 
  \label{eq:momentum}
\end{equation}
where $\tau_0$ is $\eta_2 \tau_{\rm IR}$ at the solar metallicity. 
In our fiducial model, we set $\eta_1 = 2$ since we do not explicitly 
consider momentum injection by stellar winds, the amount of which 
is the same order as that by the radiation pressure. 
We set $\tau_0 = 30$, which is consistent with the
value found by \citet{hopkins11} who reported average infrared optical
depth of $\langle \tau_{\rm IR} \rangle \sim 10-30$ in their
high resolution simulated Milky Way-like galaxy. 

During a time-step, $\Delta t$, a young star particle distributes 
the momentum and energy from radiation to surrounding gas particles. 
A gas particle thus may receive momentum, $\Delta p_{\rm rad}$, and 
energy, $\Delta E_{\rm rad}$, during any given time-step. 
The gas particle is selected to become a wind particle during 
that time-step with a probability, 
\begin{equation}
  {\cal P}_{\rm w}^{\rm rad} 
  = \frac{\Delta p_{\rm rad}}{m_{\rm SPH} v_{\rm w, rad}}, 
\end{equation}
where $v_{\rm w, rad}$ is the initial wind speed of the momentum-driven 
wind. As for the SN-driven wind, we assume that the wind speed is 
proportional to the local velocity dispersion as 
\begin{equation}
  v_{\rm w, rad} = \kappa_{\rm w}^{\rm rad} \sigma. 
\end{equation}
Since the role of the pre-SN feedback is to delay the star formation 
until the SN feedback takes place \citep{magicc}, 
we assume $\kappa_{\rm w}^{\rm rad} < \kappa_{\rm w}^{\rm SN}$ so 
that the expelled gas by the pre-SN feedback can fall back later. 
\citet{ofjt10} found that the winds with the initial wind speed 
of $3 \sigma$ are inefficient and have little effect on the 
luminosity function at $z = 0$. 
We hence employ $\kappa_{\rm w}^{\rm rad} = 3$. 

Energy conservation should impose the limit on the mass-loading of 
the momentum-driven wind. The probability, ${\cal P}_{\rm w}^{\rm rad}$, 
then becomes
\begin{equation}
  {\cal P}_{\rm w}^{\rm rad} 
  = \min\left[\frac{\Delta p_{\rm rad}}{m_{\rm SPH} v_{\rm w, rad}}, 
    \frac{\Delta E_{\rm rad}}{\onehalf m_{\rm SPH} v_{\rm w, rad}^2}
  \right]. 
  \label{eq:momentum-driven}
\end{equation}

The probability with which a gas particle is added to a wind is 
given as the sum of the two probability, 
${\cal P}_{\rm w} = {\cal P}_{\rm w}^{\rm SN} + {\cal P}_{\rm w}^{\rm rad}$,  
since they are independent events.  
We generate a uniform random number between zero and one. 
When ${\cal P}_{\rm w}$ exceeds this number, the gas particle is added to 
a wind. We determine whether the wind is SN- or radiation pressure-driven 
by drawing a uniform random number between zero and one again. 
If ${\cal P}_{\rm w}^{\rm SN} / ({\cal P}_{\rm w}^{\rm SN} +
  {\cal P}_{\rm w}^{\rm rad})$ is greater than this number, 
the gas particle is launched as an SN-driven wind with $v_{\rm w, SN}$, 
otherwise it becomes a radiation pressure-driven wind with $v_{\rm w, rad}$.  
When the value of ${\cal P}_{\rm w}$ exceeds unity, we distribute 
its excess energy and momentum to its neighboring gas particles as 
described in \citet{ofjt10}. 

\subsection{Quenching of gas cooling in large halos} 

It is now widely accepted that the stellar feedback alone cannot explain 
bright- (massive-) end of the galaxy luminosity (mass) function
and a feedback process that operates preferentially in large halos 
is needed.  (e.g. \cite{ben03}). 
High resolution X-ray observations of galaxy clusters have revealed 
large, radio-plasma cavities in intracluster media (ICM). 
These are usually associated with episodic outbursts from a central 
radio galaxy, and indicate that huge amounts of mechanical energy 
are being deposited into the ICM by powerful AGN-driven jets 
\citep{birzan04, allen06, fabian06, taylor06}. 
Simulations investigating the impact of these AGN-driven radio cavities 
suggest that this powerful feedback provides sufficient energy to offset 
the cooling radiation from the cluster and potentially explains why 
so little cool gas is seen in these systems 
\citep{quils01, churazov02, dalla-vecchia04, omma04, sijacki06}. 

Simple prescriptions for the feedback from AGN jets 
(aka. `radio-mode feedback') have been incorporated into semi-analytic 
galaxy formation models (e.g. \cite{croton06}; \cite{bower06}). 
Including radio-mode AGN feedback has resulted in dramatic improvements 
in the models' ability to match the sharp decline of the galaxy 
luminosity function and to explain the `downsizing' seen in the 
evolution of the galaxy population. 
\citet{bower06} showed that, by assuming that AGN radio-mode feedback operates 
only in quasi-hydrostatic halos where the cooling time is longer than the 
dynamical time, the galaxy luminosity functions in local and higher redshift 
universe can be matched well.

In order to model the AGN radio-mode feedback, \citet{sijacki07} and 
\citet{onb08} 
distinguished two fundamentally different modes of AGN accretion: 
radiatively efficient, geometrically thin accretion flows (standard disks: 
\cite{ss73}) and geometrically thick, radiatively inefficient accretion flows 
(which we will generically refer to as RIAFs; \cite{narayan98}; \cite{narayan05}). 
They assumed only the latter is responsible for the radio-mode feedback through 
production of powerful jets 
(e.g. \cite{rees82}; \cite{meier01}; \cite{maccarone03}; \cite{churazov05}). 
Since RIAFs exist only when the accretion rate is much lower than the Eddington 
rate \citep{narayan98}, the radio-mode feedback is naturally switched on in 
cooling inefficient large halos where central blackholes become very massive 
and gas supply to them is slowed down \citep{sijacki07, onb08}.  

We however take a simpler, phenomenological approach in this paper because 
modeling AGN feedback inevitably introduces many uncertainties, such as 
seed blackhole mass, accretion rates onto blackholes, blackhole mergers, and 
feedback from AGN. In our phenomenological treatment, we simply assume that 
the radiative cooling is suppressed in large halos where one-dimensional dark 
matter velocity dispersion is larger than $\sigma_{\rm th}(z)$.  
We parameterize the functional form of $\sigma_{\rm th}(z)$ as 
\begin{equation}
  \sigma_{\rm th}(z) = \sigma_0 (1 + z)^\alpha, 
  \label{eq:quench}
\end{equation}
where the parameter, $\alpha$, controls the redshift dependence. 
In order to reduce cooling in large halos, we modify the cooling 
function, $\Lambda(T, Z)$ as 
\begin{eqnarray}
  \Lambda(T, Z, \sigma)=\left\{ \begin{array}{ll}
      \Lambda(T, Z) & (\sigma < \sigma_{\rm th}) \\
      \Lambda(T, Z) \exp\left(-\frac{\sigma - \sigma_{\rm th}}{\beta \sigma_{\rm th}}\right) & (otherwise), \\
  \end{array} \right.
  \label{eq:agn}
\end{eqnarray}
where the parameter, $\beta$, specifies how steeply the cooling 
is suppressed above the threshold velocity dispersion, $\sigma_{\rm th}$. 
We find that a sudden suppression, i.e. $\Lambda(T, Z, \sigma) = 0$ for 
$\sigma > \sigma_{\rm th}$ creates an unwanted bump around the 
mass corresponding to $\sigma_{\rm th}$ in the galaxy stellar mass function. 
To mimic the AGN radio-mode feedback, $\alpha$ must be greater than 0 since 
the velocity dispersion of a cooling inefficient halo increases with 
increasing redshift and, for given velocity dispersion of a halo, 
the central blackhole mass is likely to become smaller with 
increasing redshift\footnote{Without the redshift dependence 
  (i.e. $\alpha = 0$), it is still possible to obtain a galaxy stellar mass 
function compatible with observational estimates at $z = 0$. We however 
lack massive galaxies at higher redshift in this case (see \cite{shimizu12}).}. 
We adopt $\alpha = 0.75$ and $\beta = 0.3$ in our fiducial model 
\citep{croton06, nagashima+05a}. 
As we will show later, this choice makes the quenching almost negligible 
at high redshift ($z > 2$) because $\sigma_{\rm th}$ becomes large and halos 
that have higher velocity dispersion than $\sigma_{\rm th}$ are quite rare. 
We note that the comparable results to this set of parameters can be 
obtained with $\sigma_0 \simeq 100$~km~s$^{-1}$ if we assume a sudden 
suppression of the gas cooling above $\sigma_{\rm th}$. 

\subsection{Models}

\begin{table*}
  \caption{Model parameters}\label{tab:params}
  \begin{center}
    \begin{tabular}{ccccccccc}
      \hline
      $c_*$ & $\eta_{\rm SN}$ & $\kappa_{\rm w}^{\rm SN}$ & $\kappa_{\rm w}^{\rm rad}$ & $\eta_1$ & $\tau_0$ & $\sigma_0$ & $\alpha$ & $\beta$\\
      & & & & & & (km~s$^{-1}$) & & \\ 
      \hline
      0.01 & 0.4 & 3.5 & 3.0 & 2.0 & 30 & 50 & 0.75 & 0.3 \\
      \hline
    \end{tabular}
  \end{center}
\end{table*}

We list the values of the model parameters in Table~\ref{tab:params}, 
which we choose to match the observational estimates of the stellar mass 
functions and the stellar mass fractions from $z = 4$ to 0. 
We show several combinations of the above feedback processes in order 
to highlight roles of each process. 
We refer to the SN feedback, the radiation pressure feedback, and the 
quenching of the cooling in large halos as `SN', `RP', and `AGN', 
respectively, for short.  
In this paper, we show the models; `SN', `SN+RP', `SN+AGN', and `SN+RP+AGN'. 
The model that employs all the feedback processes, `SN+RP+AGN', is 
our fiducial model. 
The model that only considers the SN feedback, `SN', 
corresponds to the model we used in our previous 
studies \citep{shimizu11, shimizu12, shimizu14}, although the values of the 
parameters in these studies are slightly different from those in this paper 
mainly due to the differences in the numerical resolution and cosmology. 

\section{Results} 

In this section, we compare our simulations with the available observational 
estimates from $z = 4$ to $0$ and show effects of each feedback process.   
To identify virialized dark matter halos, we first run the 
friends-of-friends (FoF) group finder \citep{defw85} with a linking 
length 0.2 in units of the mean dark matter particle separation. 
Gas and star particles near dark matter particles which compose 
a FoF group are also regarded as the member of the group. 
We then identify gravitationally bound groups of dark matter, 
gas, and stars in each FoF group by using the {\scriptsize SUBFIND} 
algorithm \citep{spr01}. 
We regard a gravitationally bound group of particles that consists 
of at least 32 particles and contains at least 10 star particles 
as a `galaxy'. 
We do not discriminate between satellite and central galaxies 
in our analyses unless otherwise stated.  
In order to exclude diffusely distributed stellar component
\citep{zibetti05, mcgee10} from the stellar component of a galaxy, 
we define the galactic stellar mass as the sum of stellar mass 
within twice the stellar half mass radius as done in 
\citet{vogelsberger13}. 

\subsection{Stellar mass}
\begin{figure*}
 \begin{center}
  \includegraphics[width=16cm]{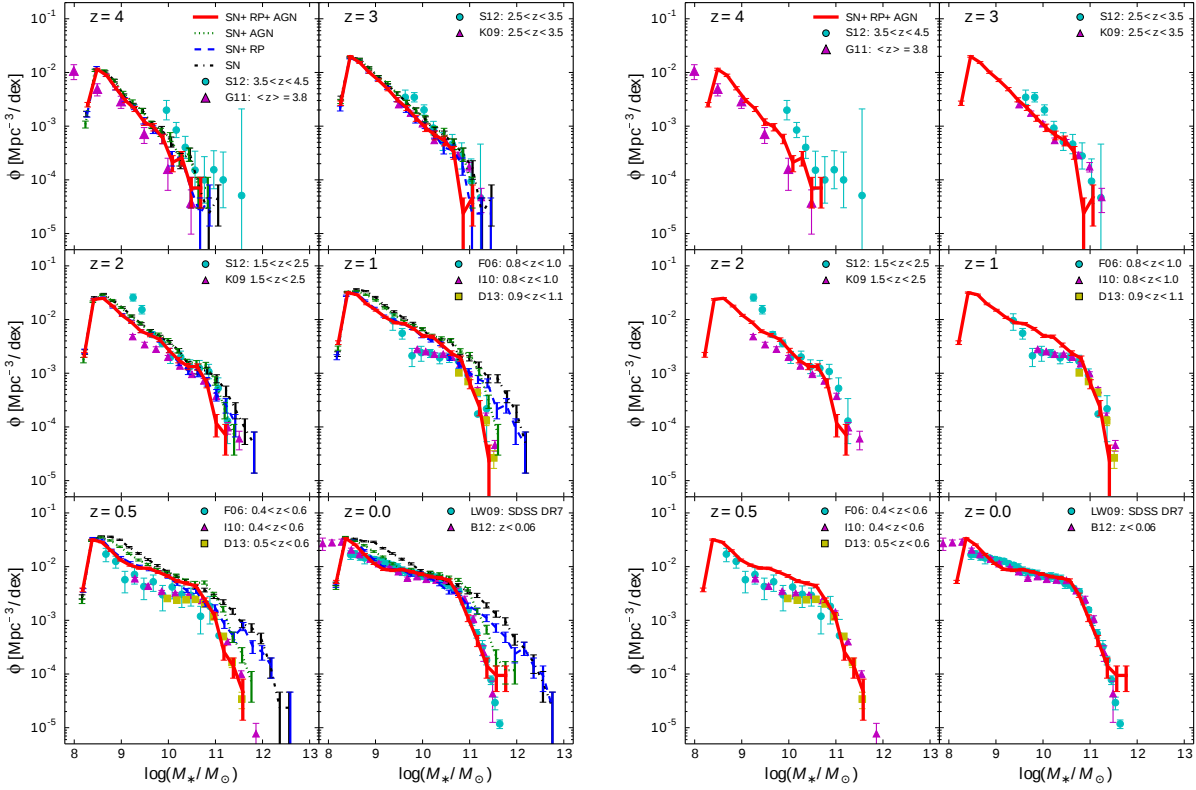} 
 \end{center}
\caption{Simulated galaxy stellar mass functions. 
  {\it Left}: The red solid, green dotted, blue dashed, and black dot-dashed
  lines illustrate the results from `SN+RP+AGN', `SN+AGN', `SN+RP', and `SN', 
  respectively. The error bars represent the $1 \sigma$ Poisson error.  
  The observational estimates at six different redshifts are also plotted as 
  the symbols with error bars. 
  The data sources are \citet{santini12} (S12) and \citet{gonzalez11} (G11) 
  for $z = 4$, S12 and \citet{kajisawa09} (K09) for $z = 3$ and 2, 
  \citet{fontana06} (F06), \citet{ilbert10} (I10), and \citet{davidzon13} 
  (D13) for $z = 1$ and 0.5, and \citet{lw09} (LW09) and \citet{baldry12} 
  (B12) for $z = 0$. {\it Right}: Same as in the left panels, but we only 
  show the fiducial model, `SN+RP+AGN', for easier comparison 
  with the observational estimates. 
}
\label{fig:smf}
\end{figure*}

Since we have chosen the model parameters so that to reproduce the 
observationally estimated stellar mass functions and stellar mass 
fractions as functions of the halo mass, 
we first show how the simulations compare to 
these estimates.  
In figure~\ref{fig:smf}, we show the simulated galaxy stellar mass functions 
at $z = 4$, 3, 2, 1, 0.5, and 0. We also plot the observational estimates 
at each redshift. 

We note that all the models agree with the observational 
estimates at $z \ge 2$. This fact may validate our previous studies 
of the high redshift galaxy populations 
\citep{shimizu11, shimizu12, shimizu14} where we only considered 
the SN-driven winds. 
At $z < 2$, however, the models without `AGN' form far too many 
massive galaxies, confirming that some quenching mechanism of 
the gas cooing in large halos is needed. 
Both `SN+AGN' and `SN+RP+AGN' reproduce the observed galaxy
stellar mass functions at all six redshifts reasonably well. 
For easier comparison with the observational estimates, 
we show the fiducial model, `SN+RP+AGN' in the right panels of 
figure~\ref{fig:smf}. 

The role of the radiation pressure-driven winds, `RP',  are not 
evident from this analysis. 
We shall investigate it in more detail by looking at the galaxy 
formation efficiency as function of the halo mass. 

\begin{figure}
  \begin{center}
  \includegraphics[width=7cm]{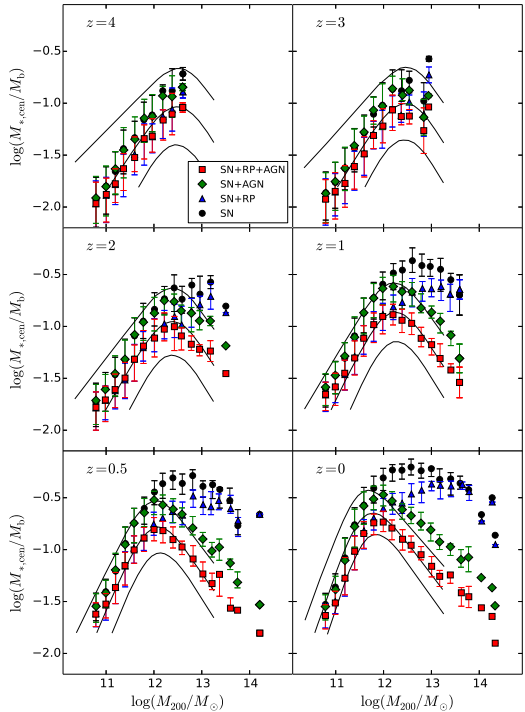} 
  \end{center}
\caption{Central galaxy formation efficiencies at $z = 4$, 3, 2, 1, 0.5, 
  and 0 as functions of the halo mass.  
  We show the median stellar mass fraction for the simulated galaxies; 
  The red squares, green diamonds, blue triangles, and black circles 
  indicate `SN+RP+AGN', `SN+AGN', `SN+RP', and `SN', respectively. 
  An error bar indicates the 25th to the 75th percentile of the
  distribution in each halo mass bin. 
  The thick black solid lines indicates the average galaxy formation 
  efficiencies estimated by \citep{moster13} using an abundance matching 
  model; the thin black lines show the $1 \sigma$ confidence level. 
}
\label{fig:moster}
\end{figure}

The technique called abundance matching places galaxies in the same 
stellar mass ranking as the dark matter halo mass rank 
\citep{conroy09, behroozi10, guo10, moster13, behroozi13}. 
Using this technique, a detailed link between observed galaxy 
stellar mass and their host halo mass has been obtained 
and it puts a strong constraint on galaxy formation models  
(see \cite{moster13}). 

In figure~\ref{fig:moster}, we compare the simulated stellar mass
fraction, $M_*/M_{200}$, for the central galaxies with that 
obtained by the abundance matching technique taken from 
\citet{moster13}, where $M_{200}$ is defined as the sum of all mass
within a sphere whose average density is 200 times the critical 
density. We only show halos that contains more than 100 dark matter 
particles since the analysis is limited to the central galaxy whose 
dark halos do not suffer from the tidal stripping. 
It is evident that the models without `AGN' convert too many 
baryons into the stars in the massive halos and cannot reproduce the 
characteristic mountain-shaped curve inferred from the abundance 
matching model. 
On the other hand, the models without `RP' predict slightly 
higher stellar fractions in low mass halos 
($M_{200} \lesssim 10^{12} M_\odot$) at 
redshift between 1 and 0.5; the pre-SN feedback brings a better 
match to the result obtained by the abundance matching model  
as suggested by the earlier studies \citep{aumer13, kannan13}. 

We note that `RP' lowers the galaxy formation efficiencies in massive 
halos more strongly than in low mass halos. 
There are two reasons why the radiation pressure feedback operates 
in large halos. Firstly, the mass-loading of the momentum-driven winds 
is proportional to $\sigma^{-1}$ which is larger than that of the 
energy-driven winds ($\propto \sigma^{-2}$) in large halos. 
Secondly, the metallicity is higher in larger halos as we will 
show later and hence more momentum per star formation is injected 
in larger halos. 
By combining all these effects, our fiducial model reproduces 
the result by the abundance matching model quite well at all 
redshifts from $z = 4$ to 0.  

We have shown that there exists a combination of the parameters 
that can reproduce the evolution of galaxy stellar mass from 
$z = 4$ to 0. 
In the following subsections, we shall examine 
other observables, such as star formation rate and metallicity, 
and the roles of each feedback process. 
We use only the galaxy stellar mass functions and 
the galaxy formation efficiencies to set our model parameters. 
The properties of the galaxies in the following subsections are 
simply {\it outcomes} of our simulations. 

\subsection{Star formation rate} 

\begin{figure}
 \begin{center}
  \includegraphics[width=7cm]{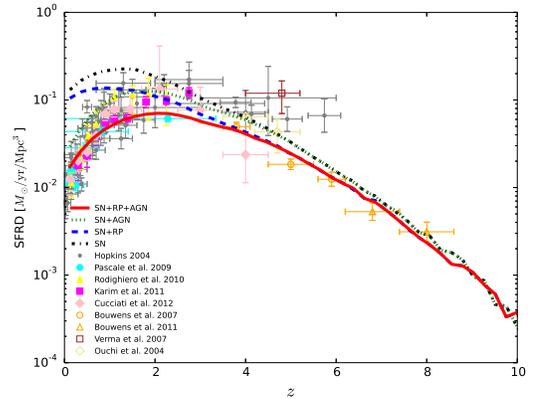} 
 \end{center}
\caption{Evolution of the cosmic star formation rate densities.  
  The red solid, green dotted, blue dashed, and black dot-dashed 
  lines illustrate the results from `SN+RP+AGN', `SN+AGN', `SN+RP', and `SN', 
  respectively.
  We also show observational constraints; the data are taken from 
  \authorcite{hopkins04} (\yearcite{hopkins04}; gray dots), 
  \authorcite{pascale09} (\yearcite{pascale09}; cyan filled circles), 
  \authorcite{rodighiero10} (\yearcite{rodighiero10}; yellow filled triangles), 
  \authorcite{karim11} (\yearcite{karim11}; magenta filled squares), 
  \authorcite{cucciati12} (\yearcite{cucciati12}; pink filled diamonds), 
  \authorcite{bouwens07} (\yearcite{bouwens07}; orange open circles), 
  \authorcite{bouwens11} (\yearcite{bouwens11}; orange open triangles), 
  \authorcite{verma07} (\yearcite{verma07}; brown open squares), and 
  \authorcite{ouchi04} (\yearcite{ouchi04}; khaki open diamonds). 
}
\label{fig:csfh}
\end{figure}

The star formation rate (SFR) tells us how the stellar component 
in the galaxies is build up and how many ionizing photons are
produced. In figure~\ref{fig:csfh}, we show the simulated cosmic star formation 
rate densities as functions of redshift. 
The models that only have the stellar feedback, i.e. `SN' and `SN+RP', 
fail to reproduce the steep decline of the SFR density toward lower 
redshift at $z < 2$. 
This suggests that the quenching of the gas cooling in large halos 
is responsible for this decline. 
By comparing the models with and without the radiation pressure feedback, 
we find that the effect of this feedback is most pronounced at the 
redshift where the SFR density is maximum ($z \simeq 2$). 
This effect makes `SN+RP+AGN' more successful in matching the galaxy 
formation efficiency to that inferred from the abundance matching 
technique at $z < 2$ (figure~\ref{fig:moster}). 
As for the cosmic SFR density, both `SN+AGN' and `SN+RP+AGN' agree with 
the observational constraints reasonably well. 

Observationally, normal star-forming galaxies lie on the so-called the 
main sequence of the star-forming galaxies, which describes the relation 
between galactic stellar mass and star formation rate 
(e.g. \cite{brinchmann04}; \cite{noeske07}; \cite{daddi07}). 
The normalization of the relationship evolves with redshift while its 
slope remains nearly constant. 
The small scatter of the relationship implies that 
most of the stellar mass of galaxies have been developed along this 
sequence \citep{noeske07}.  

\begin{figure}
 \begin{center}
  \includegraphics[width=7cm]{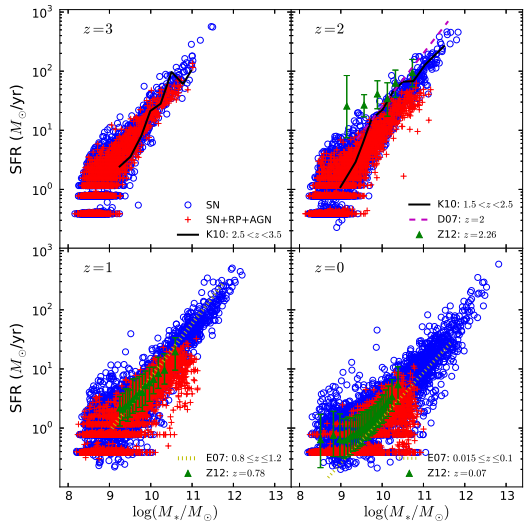} 
 \end{center}
\caption{Star formation rates against stellar masses.  
  Since the difference between the models are quite small, we 
  only show the results by `SN' (blue open circles) and `SN+RP+AGN' 
  (red plus signs). 
  The observational estimates by \authorcite{kajisawa10} 
  (\yearcite{kajisawa10}; K10), 
  \authorcite{daddi07} (\yearcite{daddi07}; D07), 
  \authorcite{elbaz07} (\yearcite{elbaz07}; E07), and 
  \authorcite{zahid12} (\yearcite{zahid12}; Z12) 
  are respectively indicated by the black solid lines, the magenta dashed 
  line, the yellow dotted lines, and the green filled triangles. 
}
\label{fig:mssfg}
\end{figure}
In figure~\ref{fig:mssfg}, we compare the simulated SFRs as 
functions of the galactic stellar mass with the several observational 
estimates at $z = 3$, 2, 1, and 0. 
For the simulations, we estimate the SFR in a galaxy from 
the mass of stars that were born in the past 100~Myr. 
Note that we plot all the galaxies that have non-zero SFRs. 
Since the difference between the models is quite small except for 
massive galaxies, we only show the most distinctive two models, 
`SN' and `SN+RP+AGN'. 

We find that our simulations agree with the observational estimates 
well. 
The slopes of the median SFRs as functions of the stellar mass 
(not shown) are $\simeq 0.9$ for $z = 3$, 2, and 1, and 
$\simeq 0.8$ for $z = 0$ also agree with the observational estimates 
(e.g. \cite{elbaz07}). 
While the simulated SFRs fall slightly below the 
observational estimates at $z = 2$, 
boosting the SFRs at this redshift would result in enhancing the 
offset between the simulated and the observationally suggested 
galaxy stellar mass functions at $z = 1$ and 0.5
(see figure~\ref{fig:smf}). 

The effect of the quenching of the gas cooling is seen even 
at $z = 3$. The absence of galaxies with very high SFRs 
($\gg 100~M_\odot~\mathrm{yr}^{-1}$), which 
exist in the SN-only simulation, may affect our predictions about 
sub-mm galaxies \citep{shimizu12}. 
We defer the investigation of this issue to future studies. 

\begin{figure}
 \begin{center}
  \includegraphics[width=7cm]{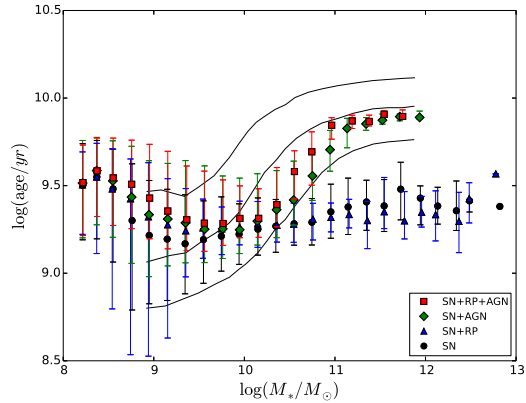} 
 \end{center}
\caption{Luminosity-weighted stellar ages as functions of 
  the stellar mass at $z = 0$.  
  We show the median $V$-band weighted stellar age for the 
  simulated galaxies. The red squares, green diamonds, 
  blue triangles, and black circles indicate `SN+RP+AGN', `SN+AGN', 
  and `SN', respectively. 
  An error bar indicates the 16th to the 84th percentile of the 
  distribution in each stellar mass bin.  
  The increase of the stellar age toward the lower stellar mass from 
  $M_* \simeq 10^{9.5}~M_\odot$ is a numerical effect 
  (see section~\ref{sec:resolution}). 
  The thick solid line indicates the observational estimate by 
  \citet{gallazzi05} and thin solid lines represent the 16th to the 84th 
  percentile of the distribution. 
}
\label{fig:age}
\end{figure}

Observationally, the galaxy stellar mass at which the SFR starts
to drop strongly decreases with time (e.g. \cite{cowie96}). 
To examine when the stellar mass of the simulated galaxies is 
built up, we show the luminosity weighted stellar ages as 
functions of the stellar mass at $z = 0$ in figure~\ref{fig:age}. 
We also show the observational estimate by \citet{gallazzi05} for 
comparison. 

Without the quenching of the gas cooling in large halos, the stellar age is 
nearly constant as a function of the stellar mass.  
The introduction of the quenching makes the stellar age 
older for $M_* \gtrsim 10^{10}~M_\odot$ and both `SN+AGN' and `SN+RP+AGN' 
broadly agree with the observational estimate. 
The inclusion of the radiation pressure feedback does not make a big 
difference, that is, `SN' and `SN+RP' are very similar to each other. 
By combined with the quenching, however, the radiation pressure 
feedback makes the stellar age older at $10^{10-11}~M_\odot$ and 
`SN+RP+AGN' better agrees with the observation than `SN+RP'. 
All the models show the increase of the stellar age toward the 
lower stellar mass from $M_* \sim 10^{9.5}~M_\odot$. This increase 
is caused by our stochastic treatment of the star formation and 
insufficient numerical resolution 
(see the resolution study in section~\ref{sec:resolution}). 
Our results suggest that the quenching of the cooling in large halos 
is essential to explain the galaxy downsizing.

\subsection{Metallicity}

We now investigate the metallicity of galaxies which is a sensitive 
diagnostic of the feedback physics. In particular, the mass-metallicity 
relation is very sensitive to the winds properties \citep{od08, ofjt10}. 

Observationally, the nebula metallicity of galaxies is determined 
by measuring a ratio (or ratios) of emission lines that are usually 
emitted from {\HII} and photo-dissociation regions.  
These lines come from the star forming regions. 
In order to compare our results with such observations, we employ 
the nebula metallicity introduced by \citet{shimizu14}, which is 
defined as the Lyman continuum weighted metallicity. 
In this definition, only the metals around the very young star particles 
are taken into account. 

\begin{figure*}
 \begin{center}
  \includegraphics[width=16cm]{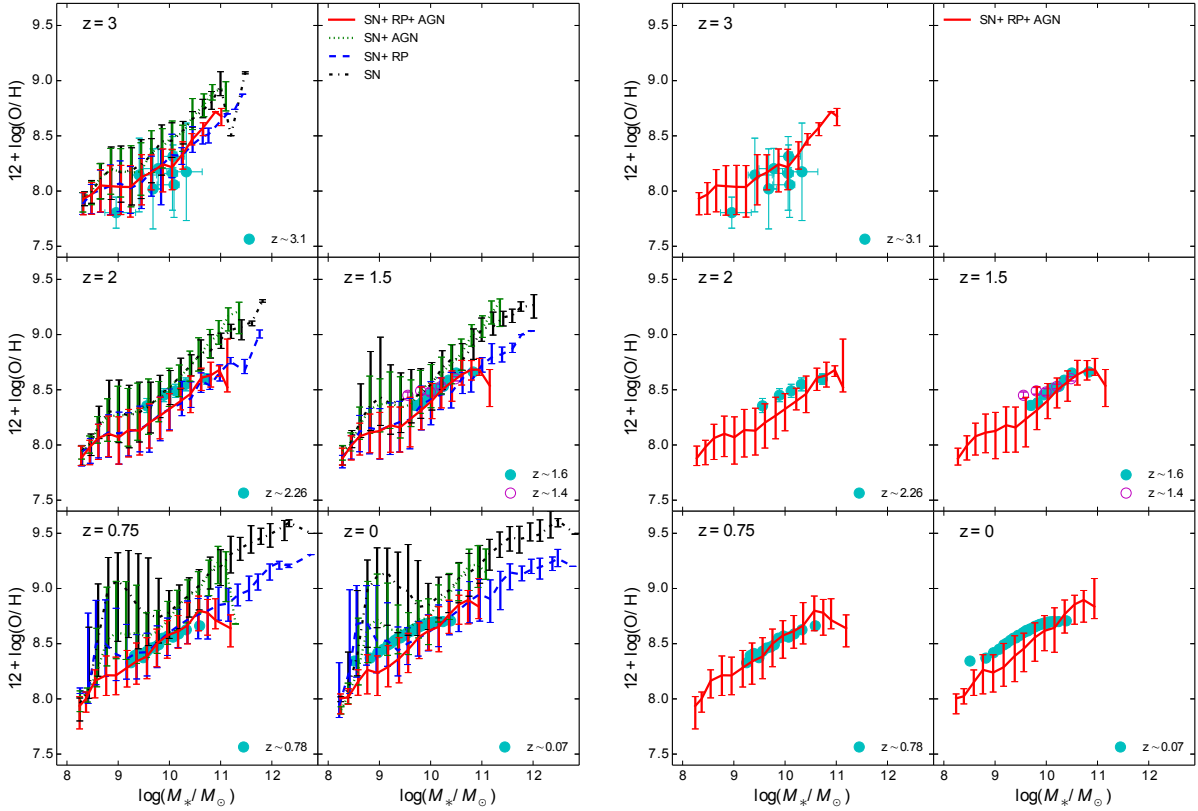} 
 \end{center}
\caption{The nebula metallicities of the simulated galaxies as 
  functions of the stellar mass at $z = 3$, 2, 1.5, 0.75, and 0. 
  {\it Left}: The binned medians for `SN+RP+AGN', `SN+AGN', 'SN+AGN', 
  and `SN' are indicated by the red solid, green dotted, blue dashed, 
  and black dash-dotted lines, respectively. 
  An error bar indicates the 25th to the 75th percentile of the 
  distribution in each mass bin. 
  Observational estimates around the presented redshifts are also shown. 
  The filled symbols in the panel for $z = 3$ are those by 
  \citet{mannucci09}, 
  The filled symbols in the panels for $z = 2$, 0.75, and 0, are the 
  observational estimates compiled by \citet{zahid12}.  
  For $z = 1.5$, estimates by \citet{zahid14} and \citet{yabe14} are 
  indicated by the filled and open symbols, respectively. 
  {\it Right}: Same as in the left panels, but we only show the fiducial 
  model, `SN+RP+AGN', for easier comparison with 
  the observational estimates. 
}
\label{fig:mzr}
\end{figure*}

In figure~\ref{fig:mzr}, we show the mass-metallicity relations at 
$z = 3$, 2, 1.5, 0.75, and 0 and compare them with the several 
observational estimates. 
There are uncertainties in defining the metallicity from a ratio (or ratios)
of emission lines \citep{nagao06, ke08}. 
We have thus converted all the data into the \authorcite{pp04} 
(\yearcite{pp04}, hereafter PP04) diagnostic.  
For the conversion between the PP04 and \citet{kobulnicky04} 
diagnostics, we apply an empirical conversion of \citet{ke08}. 
We use the relation by \citet{nagao06} to convert the metallicity 
of \citet{mannucci09} into that by the PP04 diagnostic. 
Note that because of the uncertainties in the observed metallicity 
and theoretical nucleosynthesis yields \citep{wie09b}, the normalization 
of the mass-metallicity relation is less important than its slope and 
relative redshift evolution of the relationship. 

We find that all the models show similar slopes at high redshift 
($z \ge 2$). At low redshift, however, the models without the radiation 
pressure feedback have slightly steeper profiles than those with 
the radiation pressure feedback.  
These models also show stronger 
redshift evolution than the models with the radiation pressure. 
This is because the radiation pressure feedback becomes stronger in 
higher metallicity star-forming regions (equation~(\ref{eq:momentum}))
and therefore it slows down the metal enrichment efficiently. 
Consequently, the models without the radiation pressure feedback 
show slightly too strong redshift evolution compared with the 
observations. 
Our fiducial model, `SN+RP+AGN', produces the mass-metallicity relation 
that is broadly consistent with the observational estimates, including 
the slope and the redshift evolution of the normalization because of 
the strong outward winds (see the left panels). 
We note that we would obtain slightly higher metallicity if we had 
included metal diffusion because the diffusion leads to outflowing 
particles losing metals to the circum-galactic medium 
\citep{shen10, aumer13}. 
The strange bump seen in the mass-metallicity relations  
at $M_* \simeq 10^9 M_\odot$ is due to the quantized star formation 
and insufficient resolution. 
The resolution study we will present in the next section shows that 
the bump moves to lower mass when we increase the resolution.


\section{Resolution dependence} \label{sec:resolution}

\begin{figure}
 \begin{center}
  \includegraphics[width=7cm]{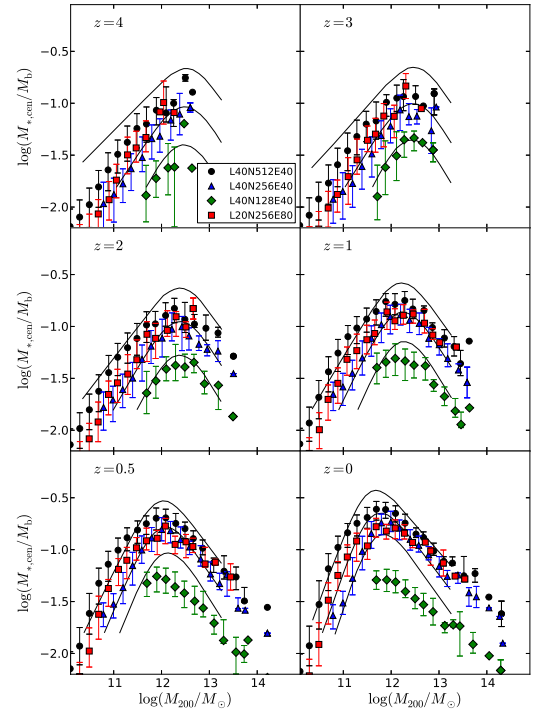} 
 \end{center}
\caption{
  Central galaxy formation efficiencies in different resolution 
  simulations. The black circles, blue triangles, and green diamonds
  represent the results from L40N512E40, L40N256E40, and L40N128E40, 
  respectively. 
  We also show the simulation with the same resolution with L40N512E40 
  but with different feedback parameters and a smaller simulation volume, 
  which is indicated by the red squares (L20N256E80).  
}
\label{fig:moster-conv}
\end{figure}
It is important to know how simulation results depend on an adopted 
numerical resolution and what parameters we should adjust when we 
change the resolution.  
For this purpose, we perform two additional simulations with the same 
parameter set as `SN+RP+AGN', to which we refer as `L40N256E40' in this 
section, but with  higher and lower numerical resolutions. 
The simulations that employ 8 times higher and 8 times lower mass 
resolutions are called `L40N512E40' and `L40N128E40', respectively. 
A gravitational softening length is a factor of 2 shorter (longer) in 
L40N512E40 (L40N128E40) than that in L40N256E40. 
We also change the star formation 
threshold density, $n_{\rm th}$, as a function of the numerical resolution; 
we adopt a density that is higher by a factor of 4 for a mass resolution 
that is higher by a factor of 8 as done in \citet{okamoto13} to improve 
the numerical convergence (see also \cite{parry12}). 

In figure~\ref{fig:moster-conv}, we compare the central galaxy formation 
efficiencies obtained from the three different resolution simulations. 
We find that our results are converging, i.e. the result from the 
intermediate resolution simulation is much closer to that from the 
high resolution one than that from the low resolution one. 
We also find that the simulations have not converged at the 
current resolution yet, i.e. the results from the high and intermediate 
resolution simulations do not agree with each other. 
In general, star formation in smaller halos is resolved with higher 
resolution and hence more stars form in a higher resolution simulation 
until a simulation resolves the smallest halo that can form stars 
after reionization 
($\sim 10^8~M_\odot$; see \cite{ogt08} and  \cite{ofjt10}).  
When a halo is more poorly resolved, winds may more easily escape from the 
halo to the intergalactic space. These two factors explain the resolution 
dependence seen in figure~\ref{fig:moster-conv}.

Since the offset is observed mainly for the low mass halos, 
we may obtain a comparable result to the fiducial simulation (L40N256E40) 
by enhancing the stellar feedback in the high resolution simulation. 
To test this, we also show the result from the simulation whose resolution 
is the same as L40N256E40 but with different values of the feedback 
parameters. 
In this simulation we increase the SN feedback efficiency from 
$\eta_{\rm SN} = 0.4$ to 0.8 and the value of $\tau_0$ in the radiation 
pressure feedback from 30 to 40. 
Other parameters are remain fixed to the fiducial values. 
In order to save the computational time, we use a smaller simulation box size, 
20~$h^{-1}$~Mpc, for this simulation. 
We call this additional simulation `L20N256E80'. 
The result from this simulation is also shown in figure~\ref{fig:moster-conv}, 
which is comparable to the fiducial simulation and consistent with 
the estimate by the abundance matching technique. 
We have confirmed that the same conclusion can be drawn from the galaxy 
stellar mass function. 

\begin{figure}
 \begin{center}
  \includegraphics[width=7cm]{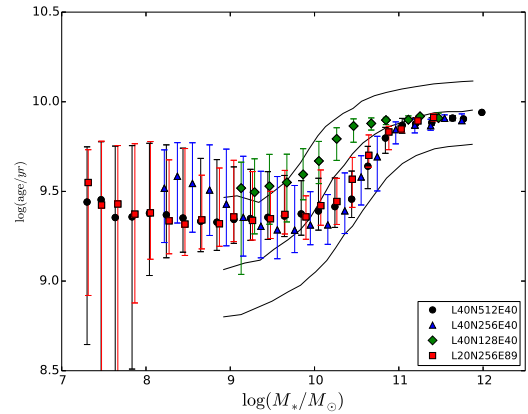} 
 \end{center}
\caption{
  Luminosity-weighted stellar ages in different resolution 
  simulations. 
  The black circles, blue triangles, and green diamonds
  represent the results from L40N512E40, L40N256E40, and L40N128E40, 
  respectively. 
  We also show the simulation with the same resolution with L40N512E40 
  but with different feedback parameters and a smaller simulation volume, 
  which is indicated by the red squares (L20N256E80).  
}
\label{fig:age-conv}
\end{figure}
Next we show resolution effects on the stellar age. 
In figure~\ref{fig:age-conv}, we compare the stellar ages as functions 
of the stellar mass in the three different resolution simulations. 
The low resolution simulation exhibits rather old stellar age 
for all the mass range. 
The results from the high and intermediate resolution simulations 
agree with each other well, indicating the results almost converge 
for the stellar age at the intermediate resolution. 
By comparing the three simulations, we conclude the upturn in the 
stellar age toward lower stellar mass is purely a numerical effect
since it occurs at much lower mass in the high resolution simulation. 

We also show the high resolution simulation with stronger 
feedback (L20N256E80). 
Interestingly, the result is almost identical to that from the
high resolution simulation (L40N512E40) in spite of the difference 
in the galaxy formation efficiency (see figure~\ref{fig:moster-conv}). 
We note that the stellar age is nearly constant below $10^{10}~M_\odot$ 
in both of the two high resolution simulations, 
where the observationally estimated stellar age gradually decreases 
toward lower mass. 
This discrepancy seems to indicate that our galaxy formation model is still 
too simple and that processes which make the star formation time-scale 
longer in smaller galaxy are required, although the simulation results are 
still broadly consistent with the observational estimate at 
$10^9 \lesssim M_* \lesssim 10^{10}~M_\odot$.

\begin{figure*}
 \begin{center}
  \includegraphics[width=16cm]{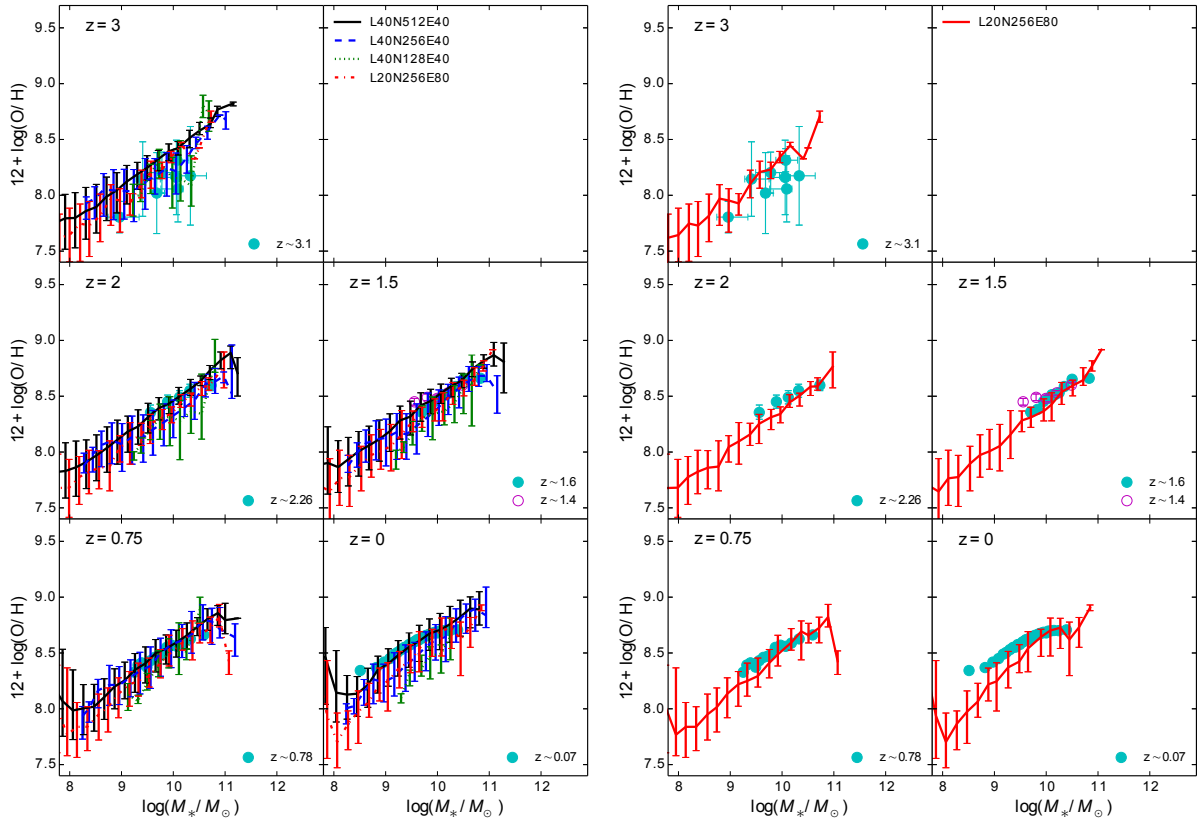} 
 \end{center}
\caption{
  The nebula metallicity of the galaxies in different resolution 
  simulations. {\it Left}:   
  The black solid, blue dashed, green dotted lines represent the results 
  from L40N512E40, L40N256E80, and L40128E40, respectively. 
  The red-dotted lines indicate the simulation with the same resolution as 
  L40N512E40 but with different feedback parameters and a smaller box size 
  (L20N256E80). 
  {\it Right}: Same as in the left panels but we only show L20N256E80 by 
  the red solid lines for easier comparison with the observational 
  estimates. 
}
\label{fig:mzr-conv}
\end{figure*}

Finally we investigate resolution effects on the mass-metallicity relation. 
In figure~\ref{fig:mzr-conv}, we show the nebula metallicities as functions 
of the stellar mass in the three different resolution simulations. 
We find that the slopes well converge  
and that all the simulations show similar redshift evolution. 
We also find that the bump seen at low-mass end is purely 
a resolution effect. 
The bump moves to lower mass when we increase the resolution. 
A higher resolution simulation exhibits higher metallicity indicating 
that the feedback effect is weaker in a higher resolution simulation. 
The high resolution simulation with stronger feedback (L20N256E80) 
recovers the result from the fiducial simulation (L40N256E40). 
The result from L20N256E80 is highlighted in the right panels, which 
shows that it is, in fact, very similar to that from L40N256E40 shown 
in the right panels of figure~\ref{fig:mzr}. 

\section{Summary and discussion}

We have updated the galaxy formation model described by \citet{ofjt10} 
to study evolution of galaxy population. 
The new model employs suppression of gas cooling in large halos 
and momentum injection by radiation pressure from massive stars in 
addition to SN feedback. 
We normalize the model parameters so that the fiducial simulation 
matches the observationally estimated stellar mass functions and 
galaxy formation efficiencies from $z = 4$ to 0. 
Interestingly, the fiducial model also well reproduces 
other observational properties of galaxy population: 
the cosmic star formation rate density, the main sequence of star-forming 
galaxies, the stellar age-stellar mass relation, and the mass-metallicity 
relation over a wide range of redshift. 

In order to investigate the roles of individual feedback processes, 
we have performed the simulations by switching on and off 
each of them alternately. 
The suppression of the gas cooling in large halo, which we view as a 
phenomenological treatment of the radio mode AGN feedback, is necessarily 
to explain  the high mass-end of the galaxy stellar mass functions and 
the decline of the cosmic star formation rate density at low redshift. 
Our simple quenching model well describes the star formation properties 
in massive galaxies. This process makes the stellar age of 
massive galaxies older and nicely reproduces the galaxy downsizing 
for massive galaxies with $M_* \gtrsim 10^{10}~M_\odot$. 
On the other hand, our high resolution simulations show that 
the stellar age of the less massive galaxies is nearly 
constant for $M_* \lesssim 10^{10}~M_\odot$, while the observations 
suggest that the less massive galaxies have younger stellar age. 
This discrepancy for the low mass galaxies might suggest that 
our galaxy formation model is still too simple and we need 
processes that make effective star formation time-scale longer 
in smaller galaxies. 

When we do not consider the radiation pressure feedback, 
the redshift evolution of the mass-metallicity relation is slightly too 
strong and the slope of the relation is too steep at low redshift. 
The metallicity dependence of the radiation pressure feedback helps  
to make the simulation result broadly consistent with the observational 
estimates. 
This fact provides a strong case for the radiation pressure feedback. 

We find that the introduction of the AGN-like feedback and the radiation
pressure feedback little affects the results at high redshift ($z > 2$). 
We thus expect that our conclusions derived from the simulations of high 
redshift galaxy population \citep{shimizu11, shimizu12, shimizu14, inoue14}, 
which only take the SN feedback into account, still hold.  
However the lack of the galaxies with SFR $\gg 100~M_\odot$~yr$^{-1}$ at 
$z > 2$, which exist in `SN' model, might become a problem to account for 
the sub-mm source number counts.  
We leave this issue for future work. 

The resolution study shows that the simulation results are converging 
numerically although the perfect numerical convergence has not been 
achieved at the current resolution. 
We show that, from a high resolution simulation, consistent results 
with the intermediate resolution simulation can be obtained 
by applying slightly stronger stellar feedback while the AGN-like 
feedback is remain unchanged. 
We expect that we would need little change in the parameter values 
if we employed even higher resolution because the simulation results 
in \citet{ofjt10} and \citet{okamoto13} nicely converged with  
$m_{\rm SPH}^{\rm orig} \lesssim 10^7~M_\odot$ for Milky Way-sized 
galaxies and with 
$m_{\rm SPH}^{\rm orig} \lesssim 10^6~M_\odot$ even for the 
Local Group satellite galaxies. 

The new model can apply for wider ranges of redshift and mass than 
the previous model that forms too massive galaxies in large halos, 
in particular, at low redshift. 
This simple model is well suited for simulations that relate high 
redshift galaxy population to the local one and for studies of coevolution 
of cluster galaxies and an ICM.

\bigskip
We would like to thank Ryu Makiya and Masahiro Nagashima for 
helpful discussion. 
Numerical simulations were carried out with Cray XC30 in CfCA at NAOJ 
and T2K-Tsukuba in Center for Computational Sciences at University 
of Tsukuba.
TO acknowledges the financial support of Japan Society for the Promotion of 
Science (JSPS) Grant-in-Aid for Young Scientists (B: 24740112). 
IS acknowledges the financial support of JSPS Grant-in-Aid for Young 
Scientists (A: 23684010). 
YN acknowledges the financial support of JSPS Grant-in-Aid for Scientific 
Research (25287050) and the FIRST program Subaru Measurements 
of Images and Redshifts (SuMIRe) by the Council for Science and 
Technology Policy. 



\end{document}